\documentclass[sigconf,anonymous=false, nonacm]{acmart}

\usepackage{booktabs} % For formal tables
\usepackage{tabularx}

\setcopyright{acmcopyright}
\acmDOI{xx.xxx/xxx_x}

\acmISBN{979-8-4007-0629-5/25/03}

\acmConference[SAC'25]{ACM SAC Conference}{March 31 –April 4, 2025}{Sicily, Italy}
\acmYear{2025}
\copyrightyear{2025}

\acmArticle{4}
\acmPrice{15.00}

\begin{document}
\title{From Theory to Practice: Demonstrators of FAIR Data Spaces Across Different Sectors}
%\titlenote{Produces the permission block, and
%  copyright information}
%\subtitle{Extended Abstract}
  
\renewcommand{\shorttitle}{SIG Proceedings Paper in LaTeX Format}

\author{Nikolaus Glombiewski$^{1}$ \quad Zeyd Boukhers$^{2,3}$ \quad Christian Beilschmidt$^{4}$ \quad Johannes Drönner$^{4}$ \quad Michael Mattig$^{4}$ \quad Artur Piet$^{5}$ \quad Robert Pietrzynski$^{5}$ \quad Mehrshad Jaberansary$^{3}$ \quad Macedo Maia$^{6}$\quad Sebastian Beyvers$^{7}$ \quad Yeliz Üçer Yediel$^{2}$ \quad Muhammad Hamza Akhtar$^{2,9}$  \quad Heiner Oberkampf$^{8}$ \quad Jonathan Hartman$^{9}$ \quad Bernhard Seeger$^{1}$ \quad Christoph Lange$^{2,9}$}
\email{Email: christoph.lange-bever@fit.fraunhofer.de}
\affiliation{$^{1}$Philipps University of Marburg\\
            $^{2}$Fraunhofer Institute for Applied Information Technology FIT\\
            $^{3}$University Hospital of Cologne\\
            $^{4}$Geo Engine GmbH\\
            $^{5}$expandAI GmbH\\
            $^{6}$Leipzig University\\
            $^{7}$Justus Liebig University Giessen\\
            $^{8}$ACCURIDS GmbH\\
            $^{9}$RWTH Aachen University
            \country{}}

\def\authorsstandarized{Nikolaus Glombiewski, Zeyd Boukhers, Christian Beilschmidt, Johannes Drönner, Michael Mattig, Artur Piet, Robert Pietrzynski, Mehrshad Jaberansary, Macedo Maia, Sebastian Beyvers, Yeliz Üçer Yediel, Muhammad Hamza Akhtar, Heiner Oberkampf, Jonathan Hartman, Bernhard Seeger, Christoph Lange}

\renewcommand{\shortauthors}{N.\ Glombiewski et al.}

\begin{abstract}
The principles of data spaces for sovereign data exchange across trusted organizations have so far mainly been adopted in business-to-business settings, and recently scaled to cloud environments.
Meanwhile, research organizations have established distributed research data infrastructures, respecting the principle that data must be FAIR, i.e., findable, accessible, interoperable and reusable.
For mutual benefit of these two communities, the FAIR Data Spaces project aims to connect them towards the vision of a common, cloud-based data space for industry and research.
Thus, the project establishes a common legal and ethical framework, common technical building blocks, and it demonstrates the orchestration of multiple building blocks in self-contained settings addressing a diverse range of use cases in domains including health, biodiversity, and engineering.
This paper gives a summary of all demonstrators, ranging from research data infrastructures scaled to industry-ready cloud environments to work in progress on building bridges between operational business-to-business data spaces and research data infrastructures.
\end{abstract}




\maketitle

% Rough page structure (page limit is 8, wiht the option to pay for 2 more pages):
% 1. We'll need one page for all authors. Ideally, the abstract and the paper's metadata will also fit on this page.
% 2. introduction
% 3. related work / background
% 4. demonstrators:
%    ± ½ page for each demonstrator – in detail:
%    1 page (page #4) for NFDI4BioDiversity (AP4.1) including GeoEngine in open call rounds 1 and 2
%    ⅔ page (on page #5) for data quality assurance (AP4.2)
%    ⅔ page (⅓ of page #6) for distributed analytics / Personal Health Train
%    ½ page for Accurids (rest of page #6)
%    ½ page for expandAI (page #7)
%    ½ page for IndiScale / battery (page #7)
% 8. 1 page for discussion and conclusion
%    1 page for references (we won't make it with less than one page)

\section{Introduction}

% why data spaces are important 
Data spaces are an integral framework for managing, storing, and processing data across various domains. 
They are designed to enable secure and efficient data sharing among diverse entities while emphasizing interoperability and data governance. 
%In the contemporary landscape of artificial intelligence (AI), the significance of data spaces has been magnified, given their role in supplying the extensive datasets required for training sophisticated AI models. Recognizing their potential, many countries have invested substantially in developing data spaces, primarily in multiple industry sectors and public administration.
%
A promising application of the data spaces concept is the facilitation of data exchange between industry and academia. Typically, academic institutions face challenges in accessing real, industry-grade data, which are crucial for research and development. Without data spaces, industrial entities often refrain from sharing their data, as they cannot retain complete control over their data.
%or ensure its ownership. 
However, data spaces provide a secure and regulated environment that supports collaborative efforts between academia and industry, ensuring that data can be shared without compromising ownership or control. %This enhances the scope of research and innovation by providing academics with access to real-world data while maintaining industrial confidence and proprietary rights.

% why FAIR data spaces 
Research organizations have established large-scale, decentralized infrastructures for research data management, putting specific emphasis on the principle that data must be FAIR, i.e., findable, accessible, interoperable, and reusable.  
With data spaces, these infrastructures have in common that data offerings are described with semantic metadata. 
Industrial data space initiatives have become aware of the FAIR principles, but so far have taken fewer measures to fully implement them, e.g., by assigning persistent identifiers (PIDs) to data offerings.  
Such differences in approaches can also be observed in handling trust (membership in national university networks vs. technical certificates of compliance).

% What the paper is presenting 
This paper presents practical demonstrations of the FAIR Data Spaces project\footnote{https://www.nfdi.de/fair-data-spaces/} across different sectors (health, biodiversity, and engineering). It explores the application of FAIR and Gaia-X principles for sovereign data exchange in business-to-business settings and their scaling to cloud environments. The aim is to connect research organizations and business settings to establish a common legal and ethical framework and to demonstrate the orchestration of multiple building blocks in diverse use cases. After introducing background in Section~\ref{sec:related}, Section~\ref{sec:demo} outlines all demonstrators.
Section~\ref{sec:conclusion} provides a conclusion.

\section{Background and Related Work}
\label{sec:related}

The theoretical foundations of FAIR Data Spaces are rooted in the FAIR Data Principles, which emphasize the need for data to be Findable, Accessible, Interoperable, and Reusable~\cite{wilkinson2016fair}. As discussed widely in the literature, Semantic Interoperability is a key aspect in data spaces because it ensures that data can be seamlessly integrated, understood, and utilized across diverse systems and domains.~\cite{boukhers2023enhancing} Implementing these principles in data spaces has been covered by many studies~\cite{hellmanzik2022towards,scerri2022common}.

While numerous data spaces partially adhere to the FAIR principles, two of most relevant initiatives are the European Open Science Cloud (EOSC) and the National Research Data Infrastructure (NFDI). EOSC aims to create a trusted environment for sharing and reusing research data across Europe \cite{david2023sustainable}. NFDI systematically manages scientific and research data \cite{hartl2021nationale}.

FAIR Data Spaces aims to bridge between these domains and industrial data spaces (e.g. Mobility Data Space) that cater primarily to industry stakeholders. The technical implementation of FAIR Data Spaces integrates several key components: The Cloud-Native Architecture~\cite{lichtenthaler2023cloud} uses several microservices, the Metadata Management~\cite{conde2024fostering} ensures that data is easily discoverable and usable, the Data Governance framework ensures policies and regulatory standards, and the Identity and Access Management (IAM) systems are crucial for controlling user access. Finally Monitoring and Compliance ensures FAIR principles and correct operation.

\section{FAIR Data Spaces Demonstrators}
\label{sec:demo}

%The FAIR Data Spaces project\footnote{https://www.nfdi.de/fair-data-spaces/} aims at connecting research data infrastructures and business-to-business data spaces with common technical building blocks.
There are eight demonstrators\footnote{https://github.com/FAIR-DS4NFDI/wiki} that use the common technical building blocks of the FAIR Data Spaces project in specific domains.
Three demonstrators (PADME, RDC Connection, and Data Quality Assurance) had originated in consortia of the German National Research Data Infrastructure (NFDI) and a key contribution was to scale them to cloud infrastructures common in industry.
The remaining demonstrators originate from a two-round open call inviting companies to implement solutions demonstrating a clear benefit for both research and industry.

\textbf{Health:}
%Demonstrator 1
The first demonstrator addresses regulations and internal policies regarding data privacy, security and data protection in health institutions that often make the data transfer process between institutions complex, making cross-site data pooling unfeasible. It uses the Platform for Analytics and Distributed Machine Learning for Enterprises (PADME)\footnote{https://portal.pht.computational.bio/}, which is a distributed data analysis environment/platform based on the Personal Health Train (PHT) “algorithm to the data” paradigm, ensuring that distributed data can be analyzed without transferring raw data~\cite{botelho:22}. 
%EDC
Using the Eclipse Data Components (EDC) framework\footnote{https://projects.eclipse.org/projects/technology.edc}, 
PADME can act as a \emph{data provider} or a \emph{data consumer} for EDC connectors. As a provider, researchers can share their results in a data space via a data catalogue, where results are organized according to specific contract definitions. As a consumer, the PADME Consumer Connector queries the available contract definitions from the data catalogue before receiving the necessary credentials for data access.
Feasibility of the PADME approach was demonstrated in various use cases \cite{Maia:23, Jaberansary:23}.

%Demonstrator 2
The second demonstrator implements healthcare and pharma use cases showcasing how, e.g., clinical research and exchange of medicinal product information can be improved and accelerated based on semantic and secure data sharing infrastructure of Gaia-X and NFDI in combination with the ACCURIDS FAIR Data registry\footnote{https://www.accurids.com} for the management of globally unique, persistent and resolvable IDs. 
Currently, data in clinical trial is often submitted to pharma companies in the form of numerous Excel sheets through contract research organizations, lacking standardized formats, making integration and reliable identification and traceability of each data point almost impossible. ACCURIDS seamlessly connects to the federated FAIR Data Spaces infrastructure including EDC connectors and showcases capabilities that solve these issues and allow pharmaceutical companies to react swiftly to new evidence.

%Demonstrator 3
The third demonstrator from expandAI\footnote{https://www.expandai.de} deals with digital health applications (DiGA in Germany).
These applications leverage, e.g., smartphone apps and web platforms to promote health and to support therapy. The approval process for DiGA in Germany is intricate and involves several regulatory steps. 
Currently, many German companies are hesitant to incorporate Artificial Intelligence (AI) into their DiGAs due to the added complexity of the approval process.
The goal of this demonstrator is a secure and legally compliant method to collect and manage data from various wearable sensor sources, which can then be used to train AI algorithms. The prototype focuses on integrating AI-based DiGAs into the healthcare ecosystem by adhering to the FAIR and Gaia-X principles. At the core of this project is a data integration center, which gathers and standardizes Parkinson patient data for each individual. AI analytics within the center enable near real-time data insights.

\begin{figure}[bt]
    \centering
    \includegraphics[width=\linewidth]{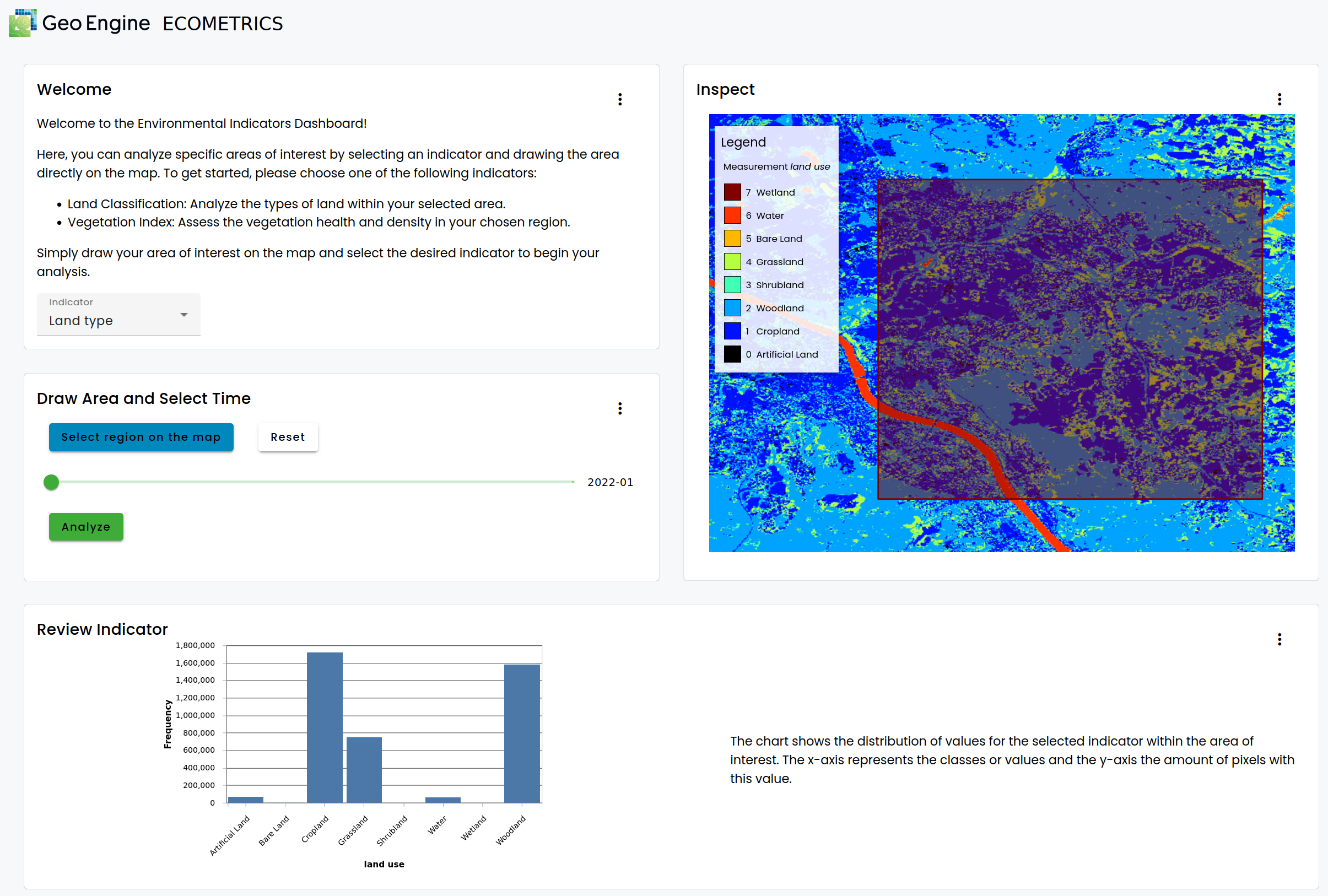}
    \caption{Dashboard of the Geo Engine-based demonstrator.}
    \label{fig:ecometrics}
\end{figure}

\textbf{Biodiversity:}
%Motivation
Identifying the loss of biodiversity is an important task for research initiatives and for industry as the European Union requires all companies to report on biodiversity by 2030\footnote{https://dfge.de/esrs-e4-biodiversitaet-oekosysteme/}.
%Overview
Our three biodiversity demonstrators use the Geo Engine, a cloud-ready geo-spatial data processing platform~\cite{DBLP:journals/dbsk/BeilschmidtDMS23}, that allows for simplified, efficient access to various data formats in multiple data spaces.
%Demonstrator 1. Challenge, Technological Solution.
The first demonstrator connects Geo Engine to the Research Data Commons (RDC)~\cite{DBLP:conf/cordi/DiepenbroekKSGD23} of NFDI4Biodiversity through establishing a data provider for the cloud-based Aruna Object Storage~\cite{Dieckmann:23}. New data sources from Gaia-X are realized via a Gaia-X Federated Catalogue\footnote{https://gitlab.eclipse.org/eclipse/xfsc/cat}. %Feasibility of connecting these different data sources is shown in a use case for detecting tree health.
%Demonstrator 2. Challenge, Technological Solution.
The second demonstrator combines data from two data spaces, namely NFDI4Biodiversity via Aruna and the Copernicus Data Space Ecosystem\footnote{https://dataspace.copernicus.eu/}, to create a machine learning training pipeline for biodiversity indicators.
In a custom dashboard, users can select an indicator, define the area and time of interest, and generate statistics that are visualized as plots 
(\autoref{fig:ecometrics}) 
based on re-usable UI components from Geo Engine’s front-end library.
%Demonstrator 3. Challenge, Technological Solution.
The third demonstrator uses similar methods to compute ESG scores for properties of land-holders while employing the concept of a virtual data trustee that logs computations and data usage.

\textbf{Engineering:}
Most present day research is not always performed with a strong focus on data integrity and maintenance. Constant vigilance regarding data quality is an often time intensive task that is open to simple automation. 
%Demo 1
The Data Quality Assurance demonstrator is a Gaia-X compatible software unit, leveraging hardware resources of the Open Telekom Cloud to allow a research group to describe their expected data in a standardized way via the Frictionless\footnote{https://frictionlessdata.io} Python package. 
The project runs within the GitLab environment. A GitLab Runner will process each file it is provided access to, and creates web-based reports (as well as machine-readable reports), allowing users to view identified issues and summary statistics without having to load the raw datasets.  

The RuQaD Batteries demonstrator builds a connection to the BatCAT (Battery Cell Assembly Twin) dataspace.
The BatCAT Horizon Europe Consortium develops a digital twin towards a more sustainable and less expensive manufacturing processes for battery technologies.
The industrial partners of BatCAT rely on data collected in academia to validate, verify and fine-tune their own findings. 
IndiScal\footnote{https://www.indiscale.com/}, as a BatCAT consortial partner, acts as a data provider for the demonstrator and offers data for reuse by the consortium after checking data for the FAIR requirements, e.g., the presence of a license and provenance information. Data exchange is enabled via the EDC framework and LinkAhead~\cite{Fitschen2019}. 
The \textit{Data Quality Assurance} demonstrator is reused for this extended use-case, thereby showing the technical viability and investigating possible business use cases as a value-added service in the data space.

%Brandt2021 - Brandt, N., Griem, L., Herrmann, C., Schoof, E., Tosato, G., Zhao, Y., Zschumme, P. and Selzer, M., 2021. Kadi4Mat: A Research Data Infrastructure for Materials Science. Data Science Journal, 20(1), p.8. DOI: http://doi.org/10.5334/dsj-2021-008

%EC2021 - European Commission, Updating the 2020 New Industrial Strategy, SWD(2021) 352, 2021.

%Amici2024 - J. Amici et al., Adv. Energy Mater. 12: 2102785, doi:10.1002/aenm.202102785, 2022.

%Edström2022 - K. Edström et al., “Inventing the sustainable batteries of the future” (BATTERY 2030+ Roadmap), 2022.

%Hornung2024 - D. Hornung et al., “Agile Research Data Management with Open Source: LinkAhead”, ing.grid 1(2), 2024. doi: https://doi.org/10.48694/inggrid.3866

%Fitschen2019 - T. Fitschen et al., Data 4: 83; doi:10.3390/data4020083, 2019

%ELNConsortium https://github.com/TheELNConsortium/

%Horsch2022 - M. T. Horsch et al., in Proc. DAMDID 2021, pp. 166–177, doi:10.1007/978-3-031-12285-9_10, 2022.

%BatCAT - doi:10.3030/101137725
\section{Conclusion}
\label{sec:conclusion}

This paper demonstrates the applicability of FAIR Data Spaces across diverse sectors, emphasizing their viability in business-to-business settings. These demonstrations serve as a bridge that facilitates the exchange between science and business settings, under a shared legal and ethical framework, paving the way for future advancements in the field.

\begin{acks}
This research has been funded by the German Federal Ministry of Education and Research (BMBF) under the grant numbers FAIRDS \{05, 08, 10, 11, 14, 15\}.
\end{acks}

\bibliographystyle{ACM-Reference-Format}
\bibliography{sample-bibliography}

\end{document}